\documentclass[aps,pre,twocolumn,showpacs,preprintnumbers,amsmath,superscriptaddress]{revtex4}
\usepackage{amsmath} \usepackage{graphicx}
\def\xv{{\bf x}}
\begin{document}

\title{Cusps, self-organization, and absorbing states}

\author{Juan A. Bonachela}
\affiliation{ Departamento de
  Electromagnetismo y F{\'\i}sica de la Materia and \\ Instituto de
  F{\'\i}sica Te{\'o}rica y Computacional Carlos I, Facultad de
  Ciencias, Univ. de Granada, 18071 Granada, Spain} 
\author{Mikko  Alava}


\affiliation{Helsinki University of Technology, Department of Applied
  Physics, HUT-02105, Finland}

\author{Miguel A.  Mu\~noz} \affiliation{ Departamento de
  Electromagnetismo y F{\'\i}sica de la Materia and \\ Instituto de
  F{\'\i}sica Te{\'o}rica y Computacional Carlos I, Facultad de
  Ciencias, Univ. de Granada, 18071 Granada, Spain}

\begin{abstract}
  Elastic interfaces embedded in (quenched) random media exhibit
  meta-stability and stick-slip dynamics. These non-trivial dynamical
  features have been shown to be associated with cusp singularities of
  the coarse-grained disorder correlator.  Here we show that {\it
    annealed} systems with many absorbing states and a conservation
  law but {\it no quenched disorder} exhibit identical cusps.  On the
  other hand, similar non-conserved systems in the directed
  percolation class, are also shown to exhibit cusps, but of a
  different type.  These results are obtained both by a recent method
  to explicitly measure disorder correlators and by defining an
  alternative new protocol, inspired by self-organized criticality,
  which opens the door to easily accessible experimental realizations.
\end{abstract}

\date{\today}
\maketitle

\begin{figure}
\includegraphics[width=75mm]{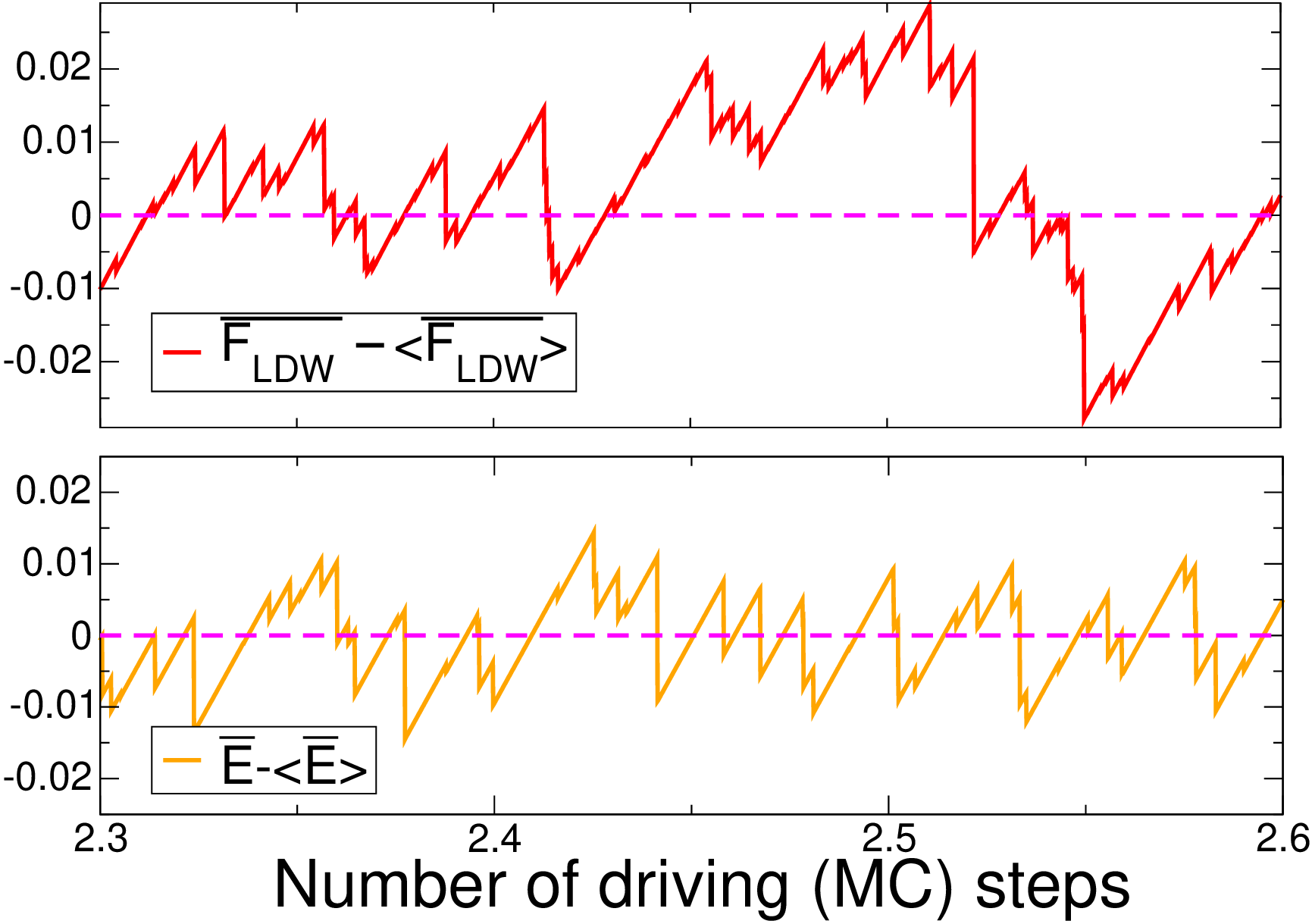}
\caption{(Color online). Up (Down): Steady state time series of the
  spatially averaged force (background field) for a single run of
  Eq.(\ref{lim}) (Eq.(\ref{FES})) using the protocol by
  Le-Doussal-Wiese (using self-organized criticality); the average
  values have been subtracted in both cases. Linear increases are
  followed by abrupt falls: the fingerprint of cusps.}
\label{landscape}
\end{figure}

Elastic objects/interfaces evolving in disordered media constitute a simple
and unified way to describe phenomena as different as wetting, dislocations,
cracks, domain walls in magnets,
or charge density waves \cite{HHZ}.  In all these examples, two conflicting
mechanisms compete: the tendency to minimize the elastic energy, leading to
flat interfaces, and the propensity to accommodate to local minima of a random
pinning potential, giving rise to rough profiles.  In this way, many different
meta-stable states exist, opening the door to the rich phenomenology
characteristic of disordered systems: slow dynamics, creeping, stick-slip
motion, avalanches, etc \cite{Balents}.  At {\it zero temperature} and in the
presence of a pulling force, a critical point separates two different phases:
i) a {\it pinned} one, in which disorder dominates and the interface remains
trapped at some local minimum of the random potential, and ii) a {\it
  depinned} phase in which the force overcomes the pinning potential and the
interface moves forward. A prominent example characterizing the dynamics of
these systems is the Quenched Edwards-Wilkinson equation:
\begin{equation}
\partial_t h(\xv,t) = \nu \nabla^2 h (\xv,t) + F + \eta(\xv,h(\xv,t))~,
\label{lim}
\end{equation}
where $\nu$ is a constant, $h(\xv,t)$ the interface height, $F$ an external
force, and $\eta(\xv,h(\xv,t))$ a quenched (random field) noise.  The noise
correlator is $ \langle \eta(\xv, 0) \eta(\xv',u) \rangle = \Delta_0(u)
\delta(\xv-\xv')$, where $\Delta_0(u)$ is some fast-decaying function. At a
critical force, $F_c$, Eq.(\ref{lim}) shows a {\em depinning transition},
representative of a broad universality class.  Analytical understanding of
Eq.(\ref{lim}) comes from various fronts, including the {\it functional
  renormalization group} (FRG) \cite{FRG,2loops}, in which the disorder
correlator, $\Delta_0(u)$, renormalizes as a function.  This approach led,
some years ago, to a successful computation of critical exponents in an
$\epsilon(=4-d)$ expansion up to one-loop order \cite{FRG}.  The emerging FRG
fixed-point function for the disorder correlator, $\Delta^*(u)$, presents the
peculiarity of having a {\it cusp singularity at the origin}, i.e.
$\Delta^{*'}(0^+) = -\Delta^{*'}(0^-) \neq 0$. Very interestingly, such
``cusps'', rather than been a curiosity, have been argued to be essential to
capture the physics of disordered systems \cite{Balents,LDW1,LDW2,Gourmet}.
Physically, the underlying idea is that the effective {\it energy landscape}
describing a random media in which an interface advances consists of a series
of parabolic wells (representing different metastable states) matching at
singular points where the first derivative (i.e. the force) is discontinuous
\cite{Balents,LDW2,Gourmet}. In this way, the total random force experienced
by the interface as it advances has necessarily a {\it sawtooth profile}, with
linear increases followed by abrupt falls (see Fig.~\ref{landscape}). These
jumps reflect the change from one metastable state to another; they dominate
the statistics of the correlation functions and generate a cusp singularity at
the origin of the renormalized disorder correlator. Le Doussal, Wiese, and
collaborators have recently extended the FRG calculation up to $2$-loops
\cite{2loops} and developed a strategy to measure disorder correlators,
allowing to test the FRG predictions in computer simulations as follows
\cite{LDW1,LDW2}. In Eq.(\ref{lim}), $F$ is replaced by a confining force $
F_{LDW}=m^2 (w-h(x,t))$, derivative of a parabolic potential centered at a
$h=w$, where $w$ is a constant.  For any $w$, a stable sample-dependent
interface configuration, with average height $\bar{h}$, is found.  Then, by
slowly increasing $w$, $F_{LDW}$ grows until, eventually, the interface
overcomes a barrier and falls into a new meta-stable state with larger
$\bar{h}$ (giving rise to sawtooth profiles).  The main breakthrough in
\cite{LDW1,LDW2} is to prove that, for a size $L$, the cumulants of
$w-\bar{h}$ can be written as:
\begin{eqnarray}
  && \langle w - \bar{h}(w) \rangle = F_c(m)/m^2, \label{cu} \\
  && \langle (w -
  \bar{h}(w)) (w' - \bar{h}(w'))\rangle_c = \Delta_m (w-w')/(m^4 L^d),
  \nonumber\\ && \langle [(w - \bar{h}(w))(w' - \bar{h}(w')) ]^3 \rangle_c =
  S_m(w-w') / (m^6 L^{2d}),\nonumber
\end{eqnarray}
and that, taking the limit $m \rightarrow 0$, $F_c(0^+)$ converges to
the critical force $F_c$ and $\Delta_{0^+}$ (resp. $S_{0^+}$)
coincides with (is a function of) the FRG fixed point, $\Delta^*(u)$
\cite{LDW2,Gourmet}. Numerical measurements of the cumulants in
Eq.(\ref{cu}) agree nicely with the $2$-loop FRG predictions
\cite{LDW2,Gourmet}.

In this paper, we show that cusps appear also in the apparently very different
realm of systems with annealed disorder; {\it systems with many absorbing
  states (AS), a conservation law, and no quenched disorder exhibit cusps
  identical to those of Eq.(\ref{lim})}.  Instead, similar models with AS in
the directed percolation class, as well as the Bak-Tang-Wiesenfeld sandpile
model have also cusps, but of different types. Moreover, we introduce an
alternative strategy to measure cusps using self-organized criticality (SOC),
i.e.  by alternating slow-driving and boundary dissipation. This provides a
practical and easy strategy to observe cusps in numerics and in experiments.

Systems with absorbing states have {\it annealed} noise and a dynamics which
leads to frozen/absorbing microscopic configurations.  Well known examples are
directed percolation (DP), the contact process, or the Domany-Kinzel automaton
\cite{HH}.  All these exhibit a transition from an absorbing to an active
phase in the very robust DP universality class, represented by the Langevin
equation: $ \displaystyle{\partial_t \rho (\xv,t)} = a \rho - b \rho^2 +
\nabla^2 \rho + \sigma \sqrt{\rho} \eta(\xv,t) $ where $a$ and $b$ are
parameters, $\rho(\xv,t)$ the activity field, and $\eta$ a zero-mean Gaussian
white noise.  The square-root noise ensures the AS condition: fluctuations
cease at $\rho=0$.  Scaling features different from DP emerge only in the
presence of extra symmetries or conservation laws \cite{HH,Romu}. Of
particular interest are systems with {\it many AS}, in which absorbing
configurations have a non-trivial structure, represented by a {\it background
  field} encoding the likeliness of absorbing configurations to propagate
activity when perturbed, and allowing for {\it metastability} to appear. Two
main classes of such systems are:

(i) The {\it Conserved-DP} (C-DP) class (which captures the gist of
stochastic sandpiles as the {\it Manna} or the {\it Oslo} one
\cite{Sandpiles}) is represented by a Langevin equation similar to
that of DP but with an extra conservation law \cite{FES,Romu,Lubeck}:
\begin{eqnarray}
  \displaystyle{\partial_t \rho (\xv,t)}& = &
  a \rho - b \rho^2 + \gamma \rho E(\xv,t) +  \nabla^2 \rho +
  \sigma \sqrt{\rho} \eta(\xv,t) \nonumber \\
  \displaystyle{ \partial_t E (\xv,t)}& = & D \nabla^2 \rho.
\label{FES}
\end{eqnarray}
\noindent
$a$, $b$, $D$ and $\gamma$ are parameters and $E(\xv,t)$ is the
(conserved and non-diffusive) background field.  Notwithstanding its
similarity with the well-behaved DP equation, Eq.(\ref{FES}) has
resisted all renormalization attempts and sound analytical predictions
are not yet available \cite{Fred}.

(ii) The {\it directed percolation class with many AS} \cite{many}.
Defined by models as the {\it pair contact process} \cite{many}, this
class has no extra symmetry/conservation-law with respect to DP. Its
corresponding Langevin equation is:
\begin{eqnarray}
  \displaystyle{\partial_t \rho (\xv,t)}& = &
  a \rho - b \rho^2 + \gamma \rho \Psi(\xv,t) +  \nabla^2 \rho +
  \sigma \sqrt{\rho} \eta(\xv,t) \nonumber \\
  \displaystyle{\partial_t \Psi (\xv,t)}& = &
  \alpha \rho - \beta \Psi \rho.
\label{PCP}
\end{eqnarray}
where $a, b, \gamma, \alpha$ and $\beta$ are parameters. Despite the
non-trivial absorbing phase, characterized by the background field
$\Psi(\xv,t)$, Eq.(\ref{PCP}) exhibits DP (bulk) criticality \cite{many}.

Remarkably, it has been conjectured that Eq.(\ref{lim}) and
Eq.(\ref{FES}) are equivalent descriptions of the same underlying
physics \cite{FES,Mapping}. In spite of the absence of rigorous proof,
there is strong theoretical and numerical evidence backing this
conjecture \cite{Mapping}. Indeed, there are heuristic arguments
which,
starting from Eq.(\ref{FES}) and defining a ``virtual interface'' $H(\xv,t)=
\int_0^t dt' \rho(\xv,t')$ which encodes the past activity history, lead to
Eq.(\ref{lim}) as an effective equation for $H(\xv,t)$.  Note that any past
noise trajectory is a frozen variable, allowing to map present-time annealed
noise into quenched disorder. Moreover, the background field $E(\xv,t)$ in
Eq.(\ref{FES}) can be identified with the force $F + \nu \nabla^2 h(\xv,t)$ in
Eq.(\ref{lim}); both encode the propensity to propagate activity/motion at
each site \cite{Mapping}.

In what follows, we {\bf a)} reproduce the results in \cite{LDW2} for
Eq.(\ref{lim}) and introduce an alternative protocol to measure them,
{\bf b)} look for cusps in Eq.(\ref{FES}) and compare them with those
in quenched systems to verify if, indeed, they both represent the same
physics, and {\bf c)} explore whether different types of cusps exist
for other AS systems.

\begin{figure}
\includegraphics[width=60mm]{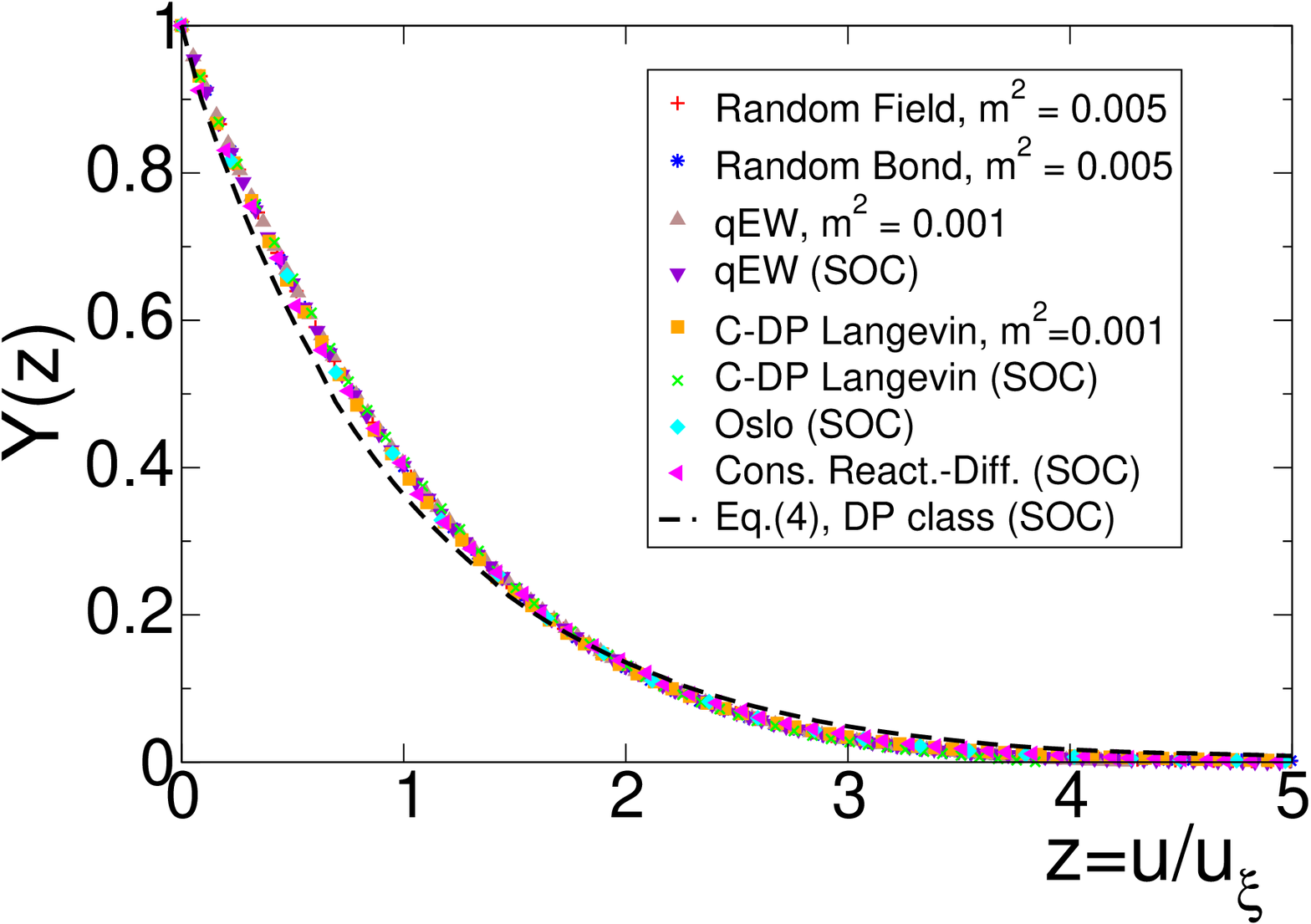}
 \includegraphics[width=60mm]{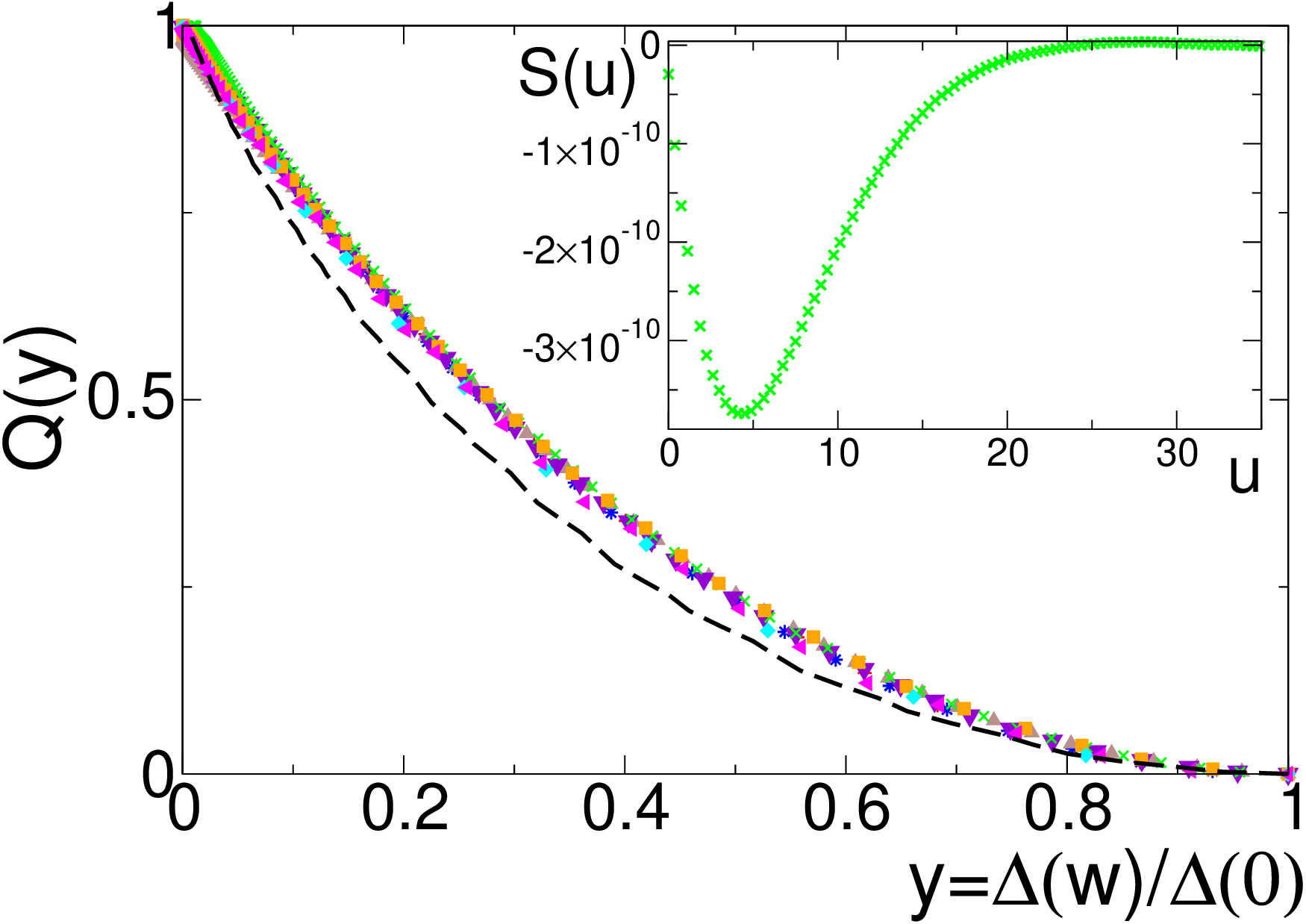}
 \caption{(Color online). Upper(Lower): $Y(z)$ ($Q(y)$) for nine
   different one-dimensional cases: i) Eq.(\ref{lim}) for random field
   and ii) random bond disorder (from \cite{LDW2}) and iii) from our
   own simulations, iv) self-organized Eq.(\ref{lim}), v) C-DP,
   Eq.(\ref{FES}), vi) self-organized Eq.(\ref{FES}), vii)
   self-organized Oslo model and, finally, viii) a critical conserved
   reaction diffusion model.  All these curves collapse into a unique
   one, different from the one obtained for ix) DP with many AS,
   Eq.(\ref{PCP}), following the self-organized procedure.
   Differences are larger for $Q(y)$ than for $Y(z)$.  Inset: function
   $S(u)$, from which $Q(y)$ is determined, for Eq.(\ref{FES}).}
\label{cusps}
\end{figure}

{\bf a)} We directly integrate a discretized version of Eq.(\ref{lim})
in one dimension, with a force $m^2 (w-h(\xv,t))$ and periodic
boundary conditions \cite{LDW2}.  $\eta(\xv,h)$ takes quenched random
values at discrete equispaced values of $h$, and is linearly
interpolated in between.  To simplify the numerics we use an overall
Heaviside step function in the r.h.s.  of Eq.(\ref{lim}) forbidding
the interface to move backwards \cite{Lesch} . For each $w$ the
dynamics eventually reaches a pinned configuration (non-positive
r.h.s. everywhere). By increasing quasi-statically the value of $w$
and using Eq.(\ref{cu}), we determine $F_c$, $\Delta(u)$ and $S(u)$.
It is convenient (see \cite{LDW1,LDW2}) to use normalized functions
defined by $Y(u/u_\xi)= \Delta(u) / \Delta(0)$ with $u_\xi$ fixed by
$\int_0^\infty Y(z) dz =1$ and $ Q (\Delta(w)/\Delta(0)) = {\int_0^w
  S(w') dw'}/ {\int_0^\infty S(w') dw'}$.  The universal functions
$Y(z)$ and $Q(y)$ are determined for up to $L=2^{12}$ and $m^2=0.001$.
Results are indistinguishable from those in \cite{LDW2} for both
functions (see Fig.~\ref{cusps}).

We now define an {\it alternative protocol} to measure cusps. The
basic idea is to let the system self-organize to its critical state by
iterating local slow driving and boundary dissipation as done in
self-organized criticality \cite{SOC,Alava}.  Starting from a pinned
interface, slow driving is implemented by increasing the force $F$
(which becomes a $\xv$-dependent function) at a randomly selected site
by a small amount. This is repeated until the interface gets depinned
at some point. Then (and only then) the dynamics, Eq.(\ref{lim}),
proceeds. 
Open boundaries $h(0)=h(L+1)=0$ allow for ``dissipation'' or
force-reduction (i.e. owing to the open-boundaries, the interface
develops, on average, a parabolic profile with {\it negative} average
elastic force) inducing the system to get pinned again, and so forth.
This driving/dissipation cycle is iterated until a steady state is
reached. Then correlations between the spatially-averaged (pulling
plus elastic) force $\overline{ F} + \overline{\nabla^2 h}$, at
different driving steps, are measured.  Note that this averaged force
defines a stochastic (generalized Ornstein-Uhlenbeck) process confined
inside a $L$-dependent potential. As the force balances at each pinned
state the corresponding random pinning force, its correlations provide
direct access to the quenched-disorder correlations. We have verified
that the average pulling force is identical to the critical force
$F_c$ (see Eq.(\ref{cu})) and measured the cumulant ratios; both
$Y(z)$ and $Q(y)$ turn out to be indistinguishable from their
counterparts above (see Fig.~\ref{cusps}).  This coincidence follows
from the fact that we are just considering two alternative
``ensembles'' to study Eq.(\ref{lim}) at criticality.  In the first,
{\it fixed velocity ensemble}, one considers closed boundaries and
lets the velocity go to $0^+$ ($m^2 \rightarrow 0$).  Instead, in the
second one, we let the system self-organize by combining slow driving
and boundary dissipation at infinitely separated timescales
\cite{SOC,Alava}. These two paths lead to the same critical properties
\cite{Narayan}.

{\bf b)} We now adapt these two protocols to study systems with many AS.  We
start analyzing Eq.(\ref{FES}) in one-dimensional lattices (size up to
$L=2^{12}$) using the recently proposed efficient integration scheme for
Langevin eqs. with square-root noise \cite{Dornic}. The system is initialized
with arbitrary initial conditions and then it self-organizes to its critical
point by means of slow driving and boundary dissipation \cite{SOC} (second
protocol) as follows: When the system is at an AS, the background and the
activity field are increased, at a randomly chosen site, by a small amount
($\delta E = \delta \rho= 0.1$); then, the dynamics proceeds according to
Eq.(\ref{FES}). Open boundaries allow to decrease the averaged background
field and hence the linear term in the activity equation, until eventually a
new AS is reached.  Then, a new driving step is performed and so on, until a
steady state is reached.  Fig.~\ref{landscape} shows the evolution of
$\overline{E}$ (minus its average value) as a function of the number of
driving steps in such a state; linear increases are followed by abrupt decays
corresponding to dissipative avalanches. From such time series, we measure the
three cumulants of the spatially averaged background field (equivalent to the
force above) and, from them, the critical point, $Y(z)$ and $Q(y)$.  Results
for $L=2^{10}$, summarized in Fig.~\ref{cusps}, show an excellent agreement
with their counterparts obtained for Eq.(\ref{lim}).

To obtain results for Eq.(\ref{FES}) using the strategy in
\cite{LDW2}, one needs to introduce a slowly moving parabolic
potential ($m^2(w-H(\xv,t))$ with $H(\xv,t) =\int_0^t dt'
\rho(\xv,t')$) for the background field and periodic boundaries.  This
is achieved by including a term $- m^2 \rho(\xv,t)$ in its Langevin
Eq.(\ref{FES}) and increasing $E$ by a small constant amount, $m^2
\delta w$, after each avalanche. When integrated in time, these two
contributions give a slowly moving parabolic potential for $E$.
Additionally, after each avalanche, activity at some random point is
created to avoid remaining trapped in the AS.  This causes a cascade
of rearrangements in the activity and background fields which
eventually stops owing to the dissipation in $E$ induced by the
negative forcing term ($-m^2 \rho$).  From the correlations of average
background field values in the AS at different steps, we obtain
results identical to those reported before for both $Y(z)$ and $Q(y)$
(see Fig.~\ref{cusps}).

The (impressive) overlap of whole functions makes a much stronger case
for the conjectured equivalence between Eq.(\ref{lim}) and
Eq.(\ref{FES}) (and also between the two protocols) than any previous
numerical agreement of critical exponents.  These conclusions have
also been extended to bidimensional systems (not shown).

Let us now study different stochastic sandpile models \cite{Sandpiles}
as well as a {\it reaction diffusion system} with conservation
\cite{Romu}, all of them argued to be in the C-DP class \cite{Alava}
but for which, owing to the discrete nature of particles/grains, the
continuous {\it statistical tilt symmetry} (which leaves the physics
of Eq.(\ref{lim}) invariant under the continuous $h(x) \to h(x) +
\delta h(x)$ transformation \cite{Gourmet}) is broken.  Using the
self-organized protocol, we produce the cusps reported in
Fig.~\ref{cusps}.  The curves overlap almost perfectly with the
previously obtained ones, implying that the continuous statistical
tilt symmetry is asymptotically restored and confirming the {\it
  universal behavior of stochastic sandpiles and their associated
  cusps} \cite{interfaces}.  Instead, for the {\it deterministic}
Bak-Tang-Wiesenfeld sandpile \cite{BTW}, known to exhibit different
critical behavior, we have measured different correlators, with
another type of cusps (not shown).

{\bf c)} Finally, we apply the previous methods to the DP (with many AS)
class, Eq.(\ref{PCP}), for which an effective interfacial description can also
be constructed \cite{DM}.  Results are summarized again in
Fig.~\ref{cusps}. $Y(z)$ shows a cusp at the origin which differs slightly
from the one above; the discrepancy in $Q(y)$ is much more pronounced,
confirming that DP exhibits a different type of scaling, and that cusps are a
common trait of systems with many AS.  This is a remarkable, so far unveiled,
property of DP systems.

In summary, we have shown the presence of identical cusps in systems
with quenched disorder, systems with AS and a conservation law, and
self-organized stochastic sandpiles. In contrast, different cusps are
obtained for systems in the directed percolation class as well as for
the deterministic Bak-Tang-Wiesenfeld sandpile.  This enriches our
present view of universality in stick-slip/avalanching systems and
confirms the ubiquitous presence of disorder correlators with cusps in
avalanching systems, with or without quenched disorder.

Our new protocol to determine cusps opens a straightforward path to measure
them in experiments.  Note that for a sandpile {\it it suffices to measure its
  weight after every avalanche to determine cusp-correlators!}  (see
\cite{ajp} for a student-laboratory set-up). Experimental studies of the mass
of granular piles have actually noticed the presence of sawtooth fluctuations
(see Fig.~1 of \cite{held}) and, for experimental ricepiles, their
correlations have been looked-at \cite{malthe}, but without determining the
relevant ones.  Similarly, cusps could be easily measured in any system, as
superconductors \cite{Wijn}, in which self-organized criticality has been
observed in the laboratory.

Challenging questions remain unanswered: Could the FRG cusps be
derived from a renormalization group solution of Eq.(\ref{FES})?
Could the DP cusps be determined analytically from Eq.(\ref{PCP})? Are
there further systems that exhibit cusps? Answering these questions
and studying experimental realizations, would greatly enhance our
understanding of stick-slip/avalanching dynamics, and would provide a
rich cross-fertilization between interfaces in random media and
systems with AS.

\vspace{0.10cm}

\end{document}